\documentclass[%
 reprint,
 amsmath,amssymb,
 aps,
 pra,
floatfix,
superscriptaddress,
raggedbottom
]{revtex4-2}

\usepackage{comment}
\usepackage{graphicx}
\usepackage{dcolumn}
\usepackage{bm}
\usepackage[normalem]{ulem}
\usepackage[
range-units = single,
]{siunitx}
\usepackage{xspace}
\usepackage{xcolor}
\usepackage[colorlinks=true, allcolors=blue]{hyperref}
\usepackage{xstring}

\makeatletter
\newcommand*{\rom}[1]{\expandafter\@slowromancap\romannumeral #1@}
\makeatother

\usepackage{xcolor}
\definecolor{lightblue}{rgb}{0., .7, 0.}

\newcommand*{\aref}[1]{%
	\IfBeginWith{#1}{eq:}{Eq.~\eqref{#1}}{}%
	\IfBeginWith{#1}{fig:}{Fig.~\ref{#1}}{}%
	\IfBeginWith{#1}{tab:}{Table~\ref{#1}}{}%
	\IfBeginWith{#1}{appendix:}{Appendix~\ref{#1}}{}%
	\IfBeginWith{#1}{sec:}{Section~\ref{#1}}{}%
}

\newcommand{\etal}{\textit{et al.\ }}
\newcommand{\be}{\begin{align}}
\newcommand{\ee}{\end{align}}
\newcommand{\ket}[1]{\ensuremath{\left| {#1} \right>}}

\newcommand{\Be}{\ensuremath{^9\mathrm{Be}^+\;}}
\newcommand{\BeNoSpace}{\ensuremath{^9\mathrm{Be}^+}}

\newcommand{\Htwop}{\ensuremath{\mathrm{H}_2^+\;}}

\newcommand{\TtwopNoSpace}{\ensuremath{\mathrm{T}_2^+}}
\newcommand{\HtwopNoSpace}{\ensuremath{\mathrm{H}_2^+}}

\newcommand{\HDpNoSpace}{\ensuremath{\mathrm{HD}^+}}

\newcommand{\Htwo}{\ensuremath{\mathrm{H}_2\;}}
\newcommand{\HtwoNoSpace}{\ensuremath{\mathrm{H}_2}}
\newcommand{\Hthreep}{\ensuremath{\mathrm{H}_3^+\;}}

\newcommand{\HtwopBe}{\ensuremath{\mathrm{H}_2^+ - {^9\mathrm{Be}}^+ \;}}

\newcommand{\HaNoSpace}{\ensuremath{\mathrm{H}}}

\newcommand{\kHz}[1]{\SI{#1}{\kilo\hertz}}
\newcommand{\MHz}[1]{\SI{#1}{\mega\hertz}}
\newcommand{\GHz}[1]{\SI{#1}{\giga\hertz}}
\newcommand{\THz}[1]{\SI{#1}{\tera\hertz}}
\newcommand{\ms}[1]{\SI{#1}{\milli\s}}

\newcommand{\mm}[1]{\SI{#1}{\milli\meter}}
\newcommand{\um}[1]{\SI{#1}{\micro\meter}}
\newcommand{\nm}[1]{\SI{#1}{\nano\meter}}

\newcommand{\EState}{\ensuremath{E,F\;^1\Sigma_g^+\;}}
\newcommand{\EStateNoSpace}{\ensuremath{E,F\;^1\Sigma_g^+}}
\newcommand{\XState}{\ensuremath{X\;^1\Sigma_g^+\;}}
\newcommand{\XStateNoSpace}{\ensuremath{X\;^1\Sigma_g^+}}
\usepackage[normalem]{ulem}

\begin{document}

\title{\textbf{State-Selective Ionization and Trapping of Single H$\mathbf{_2^+}$ Ions with (2+1) Multiphoton Ionization} 
}%

\author{Ho June Kim}
\affiliation{
Department of Physics, ETH Zürich, Zurich, Switzerland
}%
\author{Fabian Schmid}
\author{David Holzapfel}
\author{Daniel Kienzler}
 \email{daniel.kienzler@phys.ethz.ch}
\affiliation{
Department of Physics, ETH Zürich, Zurich, Switzerland
}%
\affiliation{
Quantum Center, ETH Zürich, Zurich, Switzerland
}

\date{\today}

\begin{abstract}
We report on efficient rovibrational state-selective loading of single $\mathrm{H_2^+}$ molecular ions into a cryogenic linear Paul trap using (2+1) resonance-enhanced multiphoton ionization (REMPI). The $\mathrm{H_2^+}$ ions are created by resonant two-photon excitation of $\mathrm{H_2}$ molecules from the \XState state to the \EState state, followed by nonresonant one-photon ionization. The ions are produced from \Htwo residual gas and sympathetically cooled by a co-trapped, laser-cooled \Be ion. By tuning the wavelength of the REMPI laser, we observe the loading of single \Htwop ions via the ($\nu' = 0$, $L' = 0, 1, 2$) rovibrational levels of the \EState intermediate state. We measure the success probability for the production of $\mathrm{H_2^+}$ in the ($\nu^+ = 0$, $L^+ = 1$) state via the ($\nu' = 0$, $L' = 1$) level to be  \qty{85(6)}{\percent} by quantum logic spectroscopy (QLS) of the hyperfine structure of this rovibrational state. Furthermore, we load an \Htwop ion via the ($\nu' = 0$, $L' = 2$) level and confirm its rovibrational state is ($\nu^+ = 0$, $L^+ = 2$) by QLS. We perform QLS probes on the ion over \qty{19}{\hour} and observe no decay of the rotationally excited state. Our work demonstrates an efficient state-selective loading mechanism for single-ion, high-precision spectroscopy of hydrogen molecular ions.
\end{abstract}

\maketitle

\section{Introduction}
The molecular hydrogen ion $\mathrm{H_2^+}$ and its isotopologues are of great interest for high-precision spectroscopy \cite{schiller2022precision}. Their simple three-body structure allows for precise ab initio calculations of their energy levels~\cite{korobov2017fundamental}. Combining the \textit{ab initio} calculations with high-precision spectroscopy of H$_2^+$ enables tests of fundamental physics, the determination of fundamental constants \cite{patra2020, kortunov2021proton, karr2025determination}, and the search for new physics~\cite{alighanbari2023test, delaunay2023self}. Recently, high-precision rotational and vibrational spectroscopy of \Htwop was demonstrated using molecular ensembles \cite{doran2024zero,schenkel2024laser}. However, single-ion spectroscopy holds the potential for even higher spectroscopic accuracies, enabling a more precise determination of fundamental constants and more sensitive searches for physics beyond the standard model \cite{bakalov2014electric,schiller2014simplest,karr2016hydrogen,schiller2024prospects,karr2025determination}.

One of the experimental challenges for the spectroscopy of a single \Htwop ion is rovibrational state preparation. In contrast to heteronuclear molecules such as \HDpNoSpace, the rovibrational transitions of \Htwop are dipole forbidden, and thus excited rovibrational states have lifetimes of many days~\cite{posen1983quadrupole}. We recently demonstrated quantum control of single \Htwop ions produced by electron impact ionization~\cite{holzapfel2025quantum}. To implement state preparation, helium gas was leaked into the chamber to quench the rovibrational state via the buffer gas cooling mechanism~\cite{schiller2017quantum}, limiting the accessible states to $(\nu^+ = 0, L^+ = 0, 1)$. 

Here we demonstrate resonance-enhanced multiphoton ionization (REMPI) as an alternative method for producing rovibrational state-selected \Htwop molecules in our setup. REMPI was previously implemented in other experiments for trap loading of e.g.\ hydrogen molecular ions~\cite{schmidt2020trapping, zhang2023generation}, $\mathrm{N_2^+}$ ions~\cite{tong2010, meir2019}, and O$_2^+$ ions~\cite{singh2025}. These experiments used Paul traps at room temperature and employed molecular beams which can provide high molecular densities as well as rovibrational state preparation of the neutral molecule. In contrast, we load single \Htwop ions by ionizing \Htwo molecules from the residual gas in a cryogenic Paul trap without the need for a molecular beam apparatus.

\section{(2+1) REMPI of H$\mathbf{_2}$} \label{sec:rempi}
\begin{figure}
\includegraphics[width=0.48\textwidth]{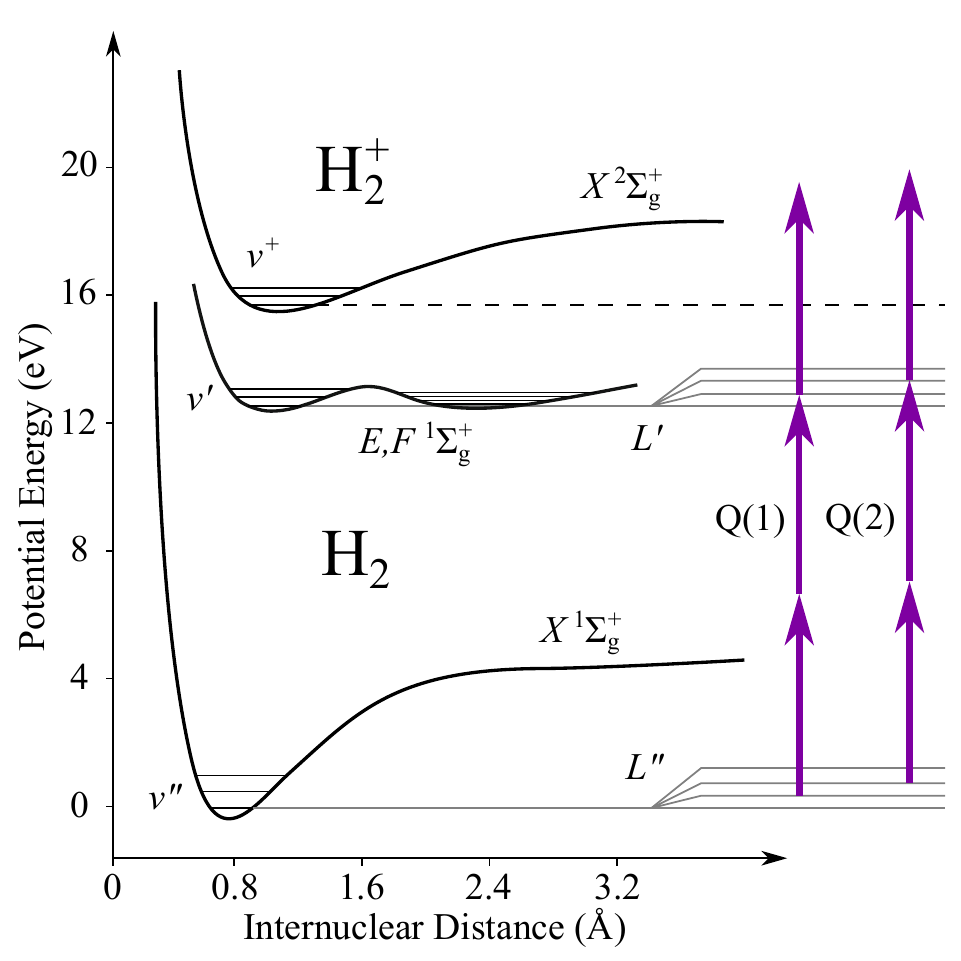}
\caption{\label{fig:rempi} \textbf{\Htwo energy-level diagram for (2+1) REMPI.} The potential-energy curves for the ground state \XStateNoSpace, intermediate state \EStateNoSpace, and ionic ground state $X\;^2\Sigma_g^+$ are shown~\cite{sharp1970potential}. Two-photon absorption brings \Htwo from the ground state to the inner well of the $E, F$ state resonantly, from which the last photon nonresonantly ionizes the molecule. Energy levels for vibrational states are drawn inside the potential-energy curves. On the right, the rotational sublevels of the vibrational ground states are shown, with examples of different two-photon rotational transitions Q(1) and Q(2). The rotational level spacings are exaggerated for visualization. Rovibrational quantum numbers for the ground state, intermediate state, and ionic ground state are given the labels $(v'', L'')$, $(v', L')$, and $(v^+, L^+)$, respectively.}
\end{figure}

We use (2+1) REMPI to produce \Htwop ions. First, \Htwo in the electronic ground state \XState with the rovibrational state ($\nu'' = 0$, $L''$) is excited to the intermediate state \EState ($\nu' = 0$, $L'$) by two-photon absorption. This is followed by nonresonant one-photon ionization resulting in a \Htwop ion in the state $X\;^2\Sigma_g^+$ ($\nu^+=0$, $L^+$)~\cite{marinero1983the,anderson1984resonance, perreault2016angular}. We use $(v'', L'')$, $(v', L')$, and $(v^+, L^+)$ to indicate the rovibrational states in the electronic ground state, intermediate state, and ionic ground state, respectively. The level diagram of the molecular states involved in (2+1) REMPI is shown in Fig.~\ref{fig:rempi}.

The rovibrational level of the intermediate state can be chosen by resolving the transitions between different rovibrational levels in the ground and intermediate electronic states.
For the expected range of the \Htwo rest gas temperature, the vibrational quantum number $\nu''$ in the electronic ground state is expected to be predominantly zero (the $\nu'' > 0$ population is \num{3e-9} at \qty{300}{\kelvin}), while the rotational quantum number $L''$ can vary.
From dipole selection rules and restrictions on rotational states due to nuclear exchange symmetry~\cite{herzberg1950}, the allowed rotational transitions for two-photon absorption between two $\Sigma$ states are O($L$), Q($L$), and S($L$) transitions, where the rotational quantum number $L$ changes by -2, 0, and +2, respectively. Our experiments concentrate on the Q transitions, which we estimate to be a factor of at least 27 stronger than the O and S transitions (see Appendix~\ref{sec:theory}).

The state-selectivity of (2+1) REMPI is compromised by the non-resonant ionization step from the \EState intermediate state. However, we expect the vibrational state to remain predominantly preserved due to a high Franck-Condon overlap between the inner well of the \EState state and the $X\;^2\Sigma_g^+$ ionic ground state~\cite{ritchie1982three}. Experimental and theoretical studies of the vibrational branching ratio for the REMPI scheme~\cite{anderson1984resonance, rudolph19872+, cornaggia1987photo} show that ionization from $\nu' = 0$ leaves the vibrational quantum number unchanged with about \qty{90}{\percent} probability. In our setup, the vibrational selectivity is further enhanced due to large photodissociation cross sections for loaded \Htwop with $\nu^+ > 0$. This means that the loading probability per pulse for $\nu^+>0$ is suppressed and the photoionization results in a loading probability of $\nu^+=0$ from $\nu'=0$ of $\sim 99.5\%$ (see Appendix~\ref{sec:simulation}).

We also expect the rotational quantum number to remain unchanged with high probability in the last ionization step. Since the bound electrons in the \EState state can be approximated with an atomic $s$ orbital, the ejected photoelectron as a result of a dipole interaction with the ionizing photon will mainly result in an $l=1$ partial wave \cite{dill1972angular, willitsch2005rovibronic}. In the creation of an $l=1$ photoelectron, the additional angular momentum of $\hbar$ coming from the ionizing photon is therefore mostly transferred to the photoelectron, and the rotation of the molecule is largely unaffected. We estimate the probability of $L^+ = L'$ to be \qty{84}{\percent} from previous studies of the photoelectron angular distribution for this REMPI process (see Appendix~\ref{sec:theory}).

\section{Experimental setup}\label{sec:exp_setup}
The experimental apparatus was previously described \cite{schwegler2023trapping,schwegler2024thesis,holzapfel2025quantum}. In this section, we summarize key aspects of the setup relevant to the current study.

\begin{figure}
\includegraphics[width=0.48\textwidth]{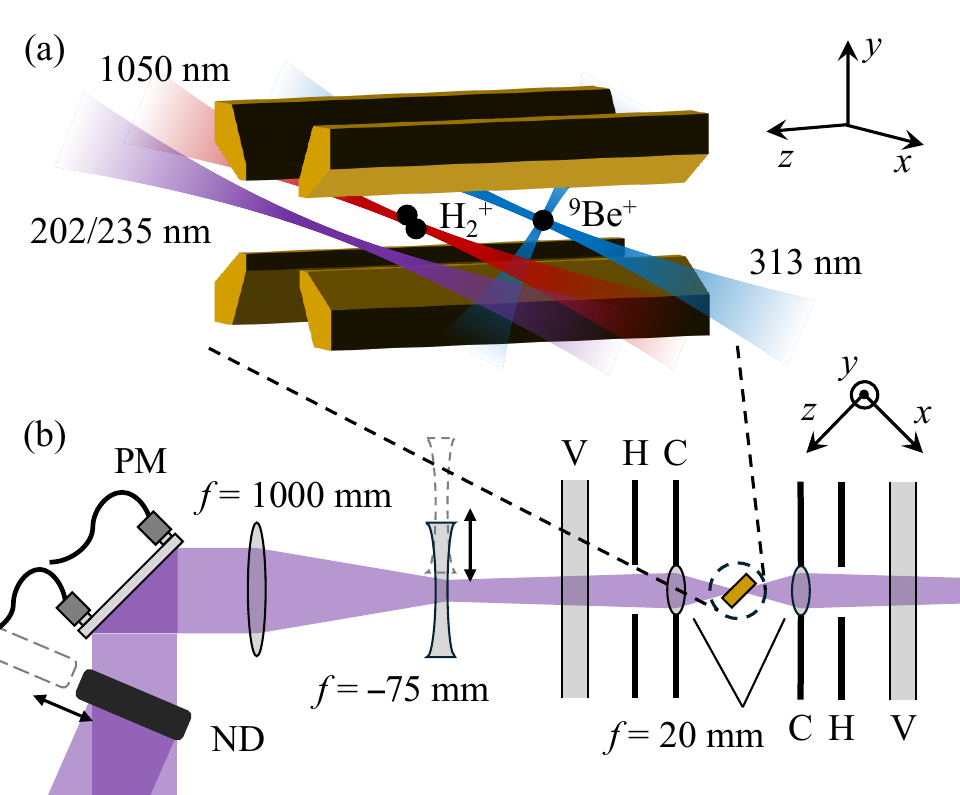}\label{fig:setup}
\caption{\textbf{Schematic of experimental setup.} (a) Oblique view of the rf electrodes and laser beams used in the setup (not to scale). The rf electrodes are \mm{2.5} long and have an ion-electrode distance of \um{300}. The dc endcap electrodes are not shown. For the chosen confinement strength the ions are \um{8.6} apart, and the beam waist radii are in the trap range from $5$ to \um{20} for the different laser beams. All laser beams propagate in the $x$-$z$ plane. The axial offset of the $202$-$\mathrm{nm}$/$235$-$\mathrm{nm}$ beam is estimated to be \um{40} from the ions' position, suppressing the dissociation of the loaded \Htwop ion (see text). The $313$ and $1050$-$\mathrm{nm}$ beams are used to control \Be and $\mathrm{H_2^+}$, respectively. (b) Top view of the REMPI beam delivery optics (not to scale). A three-lens arrangement with two lenses outside and one lens inside the vacuum chamber focus the $202$-$\mathrm{nm}$ beam at approximately the trap center. The layers of the vacuum chamber are as follows: V, outer ultra-high vacuum chamber with optical viewports; H, heat shield at $\sim\qty{75}{\kelvin}$ with apertures for optical access; and C, cryogenic inner chamber with final focusing lenses. During beryllium ionization at \nm{235}, the diverging lens is removed which compensates for the chromatic focal distance shift, and a $16$-$\mathrm{dB}$ reflective neutral density filter (ND) is inserted to attenuate the REMPI laser, since the beryllium ionization does not require as much power. A mirror on a kinematic piezoelectric mount (PM) is used to compensate for beam pointing differences between the $202$-$\mathrm{nm}$ and $235$-$\mathrm{nm}$ beams originating from the REMPI laser.}
\end{figure} 

In all experiments we load an \HtwopBe ion pair. The \Be ion is used for sympathetic cooling of the \Htwop ion and for implementing quantum logic spectroscopy \cite{schmidt2005spectroscopy,holzapfel2025quantum}. The ions are confined in a monolithic linear Paul trap. The frequency of the rf drive is \MHz{78.5}, and the motional normal mode frequencies of the \HtwopBe ion crystal are $1.3$ and \qty{3.4}{MHz} in the axial direction and $1.7$, $2.0$, $9.6$, and \qty{9.8}{MHz} in the radial directions. The electrodes and laser beams are shown schematically in Fig.~\hyperref[fig:setup]{2(a)}. The axial dc end-cap electrodes and radial rf electrodes have electrode-ion distances of $1.25$ and \mm{0.3}, respectively. Doppler cooling, state preparation, and fluorescence detection are implemented on \Be using $313$\nobreakdash-$\mathrm{nm}$ laser light. Coherent operations required for ground state cooling and for quantum logic spectroscopy (QLS) operations are implemented using Raman transitions with laser light at \nm{313} for \BeNoSpace and \nm{1050} for \HtwopNoSpace. Microwaves are used to drive magnetic dipole transitions in the hyperfine structure of both ion species.

A cryogenic chamber inside an ultrahigh-vacuum chamber houses the ion trap. The cryogenic chamber is cooled by a helium flow cryostat (Janis ST-400) supplied by a closed-cycle cryogenic system (ColdEdge Stinger). The temperature of the cold head, and thus the cryogenic chamber, can be raised from its base temperature of \qty{6}{\kelvin} by a built-in resistive heater. 

For REMPI of both Be and H$_2$ we use a tunable pulsed optical parametric oscillator (EKSPLA NT230-50-SH-DUV) which generates $3$-$\mathrm{ns}$ pulses with a $50$-$\mathrm{Hz}$ repetition rate.
To load \Be ions, neutral beryllium atoms from an oven are photoionized in a (1+1) REMPI process at \nm{235} in the trap center. To load \Htwop ions, \Htwo from the residual gas is ionized with the described (2+1) REMPI process at \qty{202}{nm}. The specified laser linewidth is less than \GHz{240} at \nm{235} and less than \GHz{150} at \nm{202}. The pulse energy at the laser output is \SI{2}{\milli\joule} at \nm{235} and \SI{570}{\micro\joule} at \nm{202}. The output beam has a flat-top profile with a beam radius of about \mm{2.5}. The $M^2$ factors for the two wavelengths were extracted by scanning beam profiler measurements and were found to be 30 for \nm{235} and 16 for \nm{202}.

The REMPI laser is delivered through the trap by the optics depicted in Fig.~\hyperref[fig:setup]{2(b)}. A three-lens arrangement with two lenses outside ($f = \mm{1000}$, $f = \mm{-75}$) and one lens inside ($f = \mm{20}$) the vacuum chamber focus the $202$\nobreakdash-$\mathrm{nm}$ beam at approximately the trap center. During beryllium ionization at \nm{235}, the diverging lens is removed to compensate for a chromatic focal distance shift. Using a duplicate optical test setup, we measure approximately Gaussian intensity profiles at the waist with a second-moment beam waist radius $w_0$ of \um{17} for \nm{202} and \um{30} for \nm{235} at the location of the trap center. We estimate the pulse energy at the trap center is \SI{200}{\micro\joule} for \nm{202} due to losses from the optical elements. Assuming a square temporal pulse shape and a Gaussian beam profile at the focus, we define the spatially averaged peak intensity from the REMPI laser for \nm{202} as $E_p/(t_p \pi w_0^2) = \SI{7}{\giga\watt/\centi\meter^2}$, where $E_p$ is the pulse energy and $t_p$ is the pulse duration.

The \Htwo number density inside the cryogenic chamber is less than $\qty{1.6(1.3)E3}{\per\centi\meter\cubed}$~\cite{schwegler2023trapping} which is too low to efficiently load \HtwopNoSpace. We thus raise the \Htwo pressure during the photoionization. We can heat a nonevaporable getter (NEG; SAES Group NEXTorr D500-StarCell) to release \HtwoNoSpace. The NEG element has a high capacity and is used to provide an elevated, approximately constant \Htwo pressure. However, the time constant for heating up and cooling down is high (\qty{30}{\min}), making it less useful as a general tool for loading. Low time constants are required to enable fast loading and also to reach a low \Htwo pressure quickly after successful loading to suppress the chemical reaction $\Htwo + \Htwop \rightarrow \Hthreep + \HaNoSpace$, which would limit the trapping lifetime \cite{schwegler2023trapping}. The reaction $\Be + \mathrm{H}_2 \rightarrow \mathrm{BeH}^+ + \mathrm{H}$ is energetically allowed only if \Be is in an electronic excited state \cite{raimondi1983spin}. Due to the low average \Be $2p$-state population ($\lesssim 0.1$), the reaction is suppressed and is less of a concern.

In QLS experiments, where long trapping lifetimes are essential, we rely on two alternative approaches to increase the \Htwo pressure, both of which achieve lower time constants than the NEG. We use either the degassing function of a hot cathode ion gauge (Kurt J. Lesker Company G8130), which releases \Htwo similar to the NEG, or we raise the cryostat temperature to between $15$ and \qty{30}{\kelvin}, evaporating \Htwo from cold surfaces and reducing the cryogenic pumping speed for \Htwo entering the cryogenic chamber. Comparing the loading rates of the three different methods for $ortho$- and $para$-\HtwopNoSpace, we observe that the loading of $ortho$-\Htwo ($L^+=1$) is suppressed when the cryostat is used for increasing the \Htwo pressure. This matches our expectation of the copper elements and the charcoal getter of the cryogenic chamber acting as \Htwo $ortho$-$para$ converters~\cite{ilisca1992ortho}.

\section{Single $\mathbf{H_2^+}$ Loading with (2+1) REMPI}
Our general experimental procedure to load a \HtwopBe ion pair is to first load a single \Be ion. A subsequently loaded \Htwop molecule will be sympathetically cooled by the \Be ion which is continuously Doppler cooled. Once sufficiently cold, the ions crystallize which leads to a shift of the \Be ion's position. During loading of \Htwop we monitor the \Be ion position in the trap by imaging its fluorescent light using a camera. Real-time analysis of the camera images allows identification of the \Be ion's shift and thus the successful loading of an \Htwop molecule. The crystallization of the \HtwopBe ion pair and the \Be position detection require $\sim\ms{500}$.

Since our \Htwop loading probability is only $\sim \qty{0.2}{\percent}$ per REMPI laser pulse, it is desirable to send REMPI laser pulses at the laser repetition rate until an \Htwop ion is loaded. This approach is complicated by the fact that the REMPI laser itself leads to photodissociation of \Htwop and that the pulse repetition interval of the laser (\qty{20}{ms}) is much shorter than the time required to detect the \Htwop loading. With a photodissociation cross section of $\sigma_d = $ \SI{1.9e-20}{cm^2} \cite{dunn1968photo, dunn1968tables} and the estimated REMPI laser intensity (see Sec. \ref{sec:exp_setup}), \Htwop is dissociated with a probability of \SI{60}{\percent} per pulse if the REMPI beam is aligned with the ion position at the trap center. Thus, continuously running the pulsed laser would likely dissociate the loaded \Htwop before it can be detected. The dissociation probability can be reduced by aligning the waist of the REMPI laser to an off-center position, relying on the fact that the ion can be ionized and captured there, but crystallizes in the trap center. In our setup, it is challenging to determine the offset of the REMPI laser waist from the trap center. We observe a dissociation probability of \SI{0.9(2)}{\percent} per pulse. Assuming a Gaussian profile, this would result in an offset of the REMPI beam from the trap center of \um{40}, but due to the poor beam shape of the REMPI beam, this can serve only as a rough estimate. To speed up the loading procedure, we use bursts of 50 REMPI laser pulses instead of single pulses, each burst interleaved by a $500$-$\mathrm{ms}$ wait time for crystallization and detection of \Htwop without dissociation. We repeat this sequence until \Htwop loading is detected.

\subsection{Spectrum and pulse energy dependence}
\begin{figure}
\includegraphics[width=0.48\textwidth]{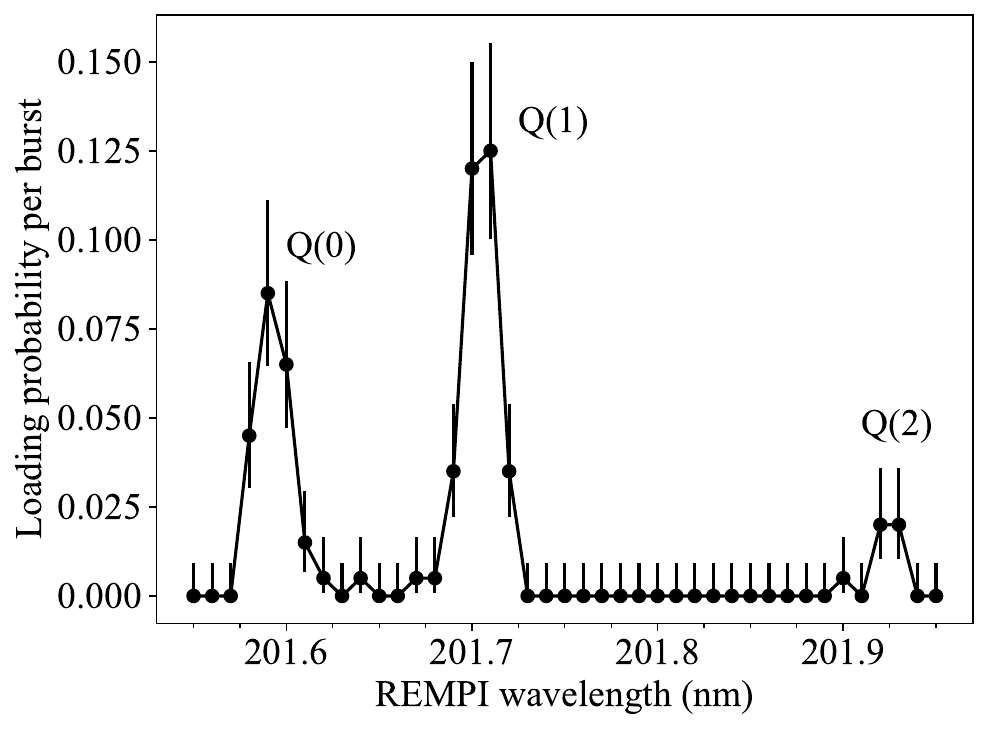}
\caption{\label{fig:two_photon} \textbf{H$\mathbf{_2^+}$ loading probability per 50-pulse burst as a function of REMPI wavelength.} The wavelength values are given in air. Error bars indicate the 68\% confidence interval for the mean number of ions assuming Poissonian statistics. Straight lines are drawn between data points to guide the eye. We can identify the resonances as the Q(0), Q(1), and Q(2) transitions between the ground \XState state and intermediate \EState state~\cite{hannemann2006frequency}. We also observed two loading events at a wavelength of \nm{202.26}, corresponding to the Q(3) transition. This single data point was measured separately after the scan and is not included in the plot.}
\end{figure}
To measure a full REMPI spectrum, we operate with a constantly elevated \Htwo pressure by heating the NEG element. 
We execute the described loading sequence and record the number of loading events for a total of 10000 pulses (200 bursts with each burst containing 50 pulses) per wavelength setting of the REMPI laser. After each loading event, we remove the \Htwop from the trap by photodissociating it with the REMPI laser tuned offresonant from any transition. The dissociation creates a hydrogen atom and a proton. We have never observed a trapped proton, which is likely due to the poor sympathetic cooling for the large mass mismatch between the proton and \Be and the high Mathieu $q$-parameter value for the proton. 

The resulting REMPI spectrum is shown in Fig.~\ref{fig:two_photon}. The locations of the resonances are in agreement with the wavelengths of the Q($L$) transitions with $L = 0, 1, 2$~\cite{hannemann2006frequency}. 
Additionally, we have observed two loading events at the predicted wavelength for Q(3) of \nm{202.26} after a total of 10000 REMPI laser pulses. No loading after 10000 pulses was observed at the predicted wavelengths for the rotation-changing S(0) and S(1) transitions which could be explained by the difference in rotational line strengths for Q and S transitions (see Appendix~\ref{sec:theory}).

\begin{figure}
\includegraphics[width=0.48\textwidth]{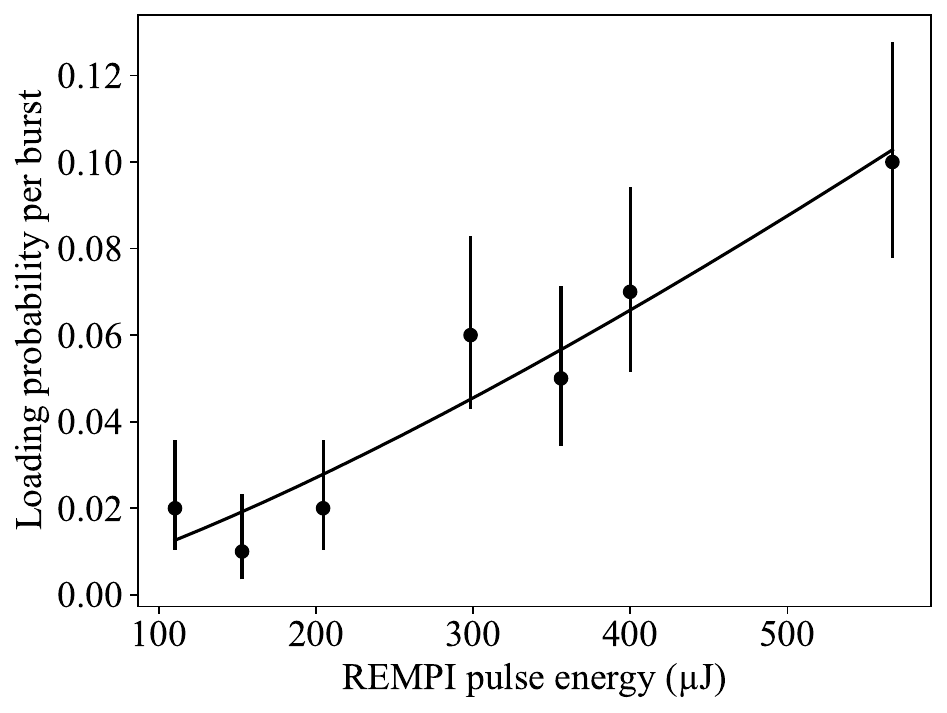}
\caption{\label{fig:power_law} \textbf{H$\mathbf{_2^+}$ loading probability per 50-pulse burst as a function of REMPI pulse energy.} The vertical error bars indicate 68\% confidence intervals for the mean number of ions assuming Poissonian loading statistics. The pulse energy error bars indicating the expected 1$\sigma$ uncertainties in pulse energy are smaller than the markers. We then fit a power-law curve to the data points using maximum likelihood estimation assuming Poissonian statistics, from which we extract an exponent of $1.28 \pm 0.31$.}
\end{figure} 

To determine the saturation of the REMPI process, we measure the \Htwop loading probability as a function of pulse energy using the Q(0) resonance (Fig.~\ref{fig:power_law}). For this experiment, we raise the cryostat temperature to elevate the \Htwo pressure.
A power-law curve $R = \alpha E^\beta$ describing the exponential dependence of the ion loading rate $R$ on the pulse energy $E$ of the REMPI laser is fitted using maximum likelihood estimation, assuming Poissonian statistics, to find the best-fit values for parameters $\alpha$ and $\beta$. We extract an exponent value of $\beta = 1.28 \pm 0.31$ from the fit. This indicates partial saturation of the three-photon process and is in good agreement with simulations of the REMPI dynamics (see Appendix \ref{sec:simulation}).

\subsection{Rotational selectivity}
We confirm the rotational state $L^+=1$ of the loaded \Htwop after REMPI by probing QLS of the hyperfine structure using the same quantum logic protocols for state preparation and readout as described in Ref.~\cite{holzapfel2025quantum}. Over 34 loading events we observe the $L^+=1$ QLS signal 29 times. Assuming no observation of the QLS signal corresponds to the creation of ions with other rotational states, we can interpret these data as an estimate of the rotational selectivity of \qty{85(6)}{\percent}, which is in agreement with the theoretical estimates from literature (given in Appendix~\ref{sec:theory}).

In our previous implementation of state preparation of \HtwopNoSpace, helium buffer gas was leaked in for rovibrational state preparation by buffer-gas cooling. Residual helium gas present in our setup from these previous experiments could have increased the apparent rovibrational selectivity by quenching $ortho$-\Htwop ions with $L^+ > 1$ to $L^+=1$. We checked that helium does not influence the rovibrational selectivity estimation with two measurements. First, collisions of residual gas with the ion pair can cause the ion order to reverse. We measured the ion order using camera images of the \Be ion for a duration of \qty{14}{\hour} and observed one reorder event, which indicates a very low collision rate that is unlikely to provide substantial rovibrational quenching. Further evidence is provided in the following section by observing the $L^+=2$ quantum logic signal.

\subsection{Quantum logic detection of $\mathbf{L^+=2}$}
\begin{figure}
\includegraphics[width=0.48\textwidth]{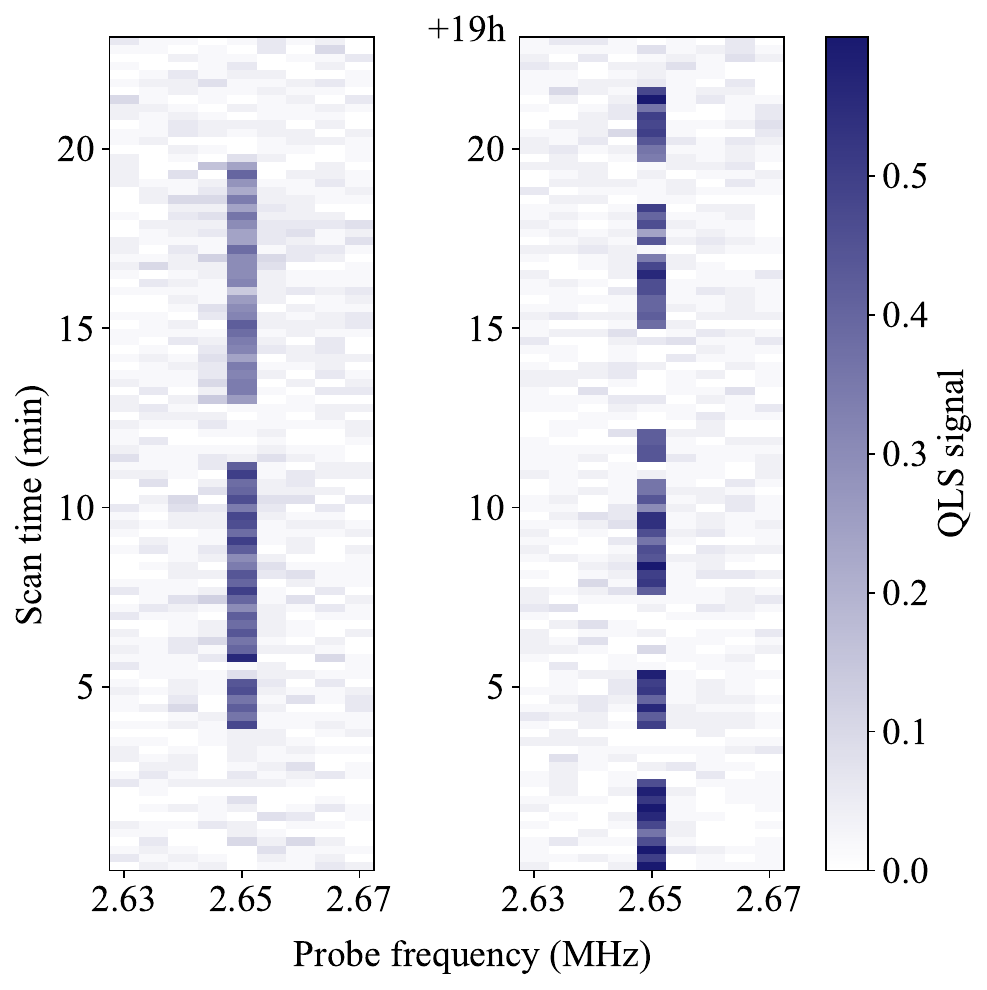}
\caption{\label{fig:qls} \textbf{QLS signal from the spin-rotation structure of H$\mathbf{_2^+}$ in the ($\mathbf{\nu^+ = 0, L^+ = 2}$) state.} After loading a \Htwop molecule using the Q(2) resonance, we confirm the rotational state by probing several transitions in the spin-rotation and Zeeman structure of \Htwop by QLS. The observed transition frequencies are in agreement with theoretical predictions for $L^+ = 2$. The plots show two example scans of the same transitions over time, scanning the frequency of the QLS probe around the expected resonance. The left plot is shortly after \Htwop loading, the right plot is \qty{19}{\hour} later. The QLS signal appears and disappears randomly due to changes in the spin-rotation-Zeeman state, likely caused by the QLS operations and/or the ac magnetic field of the Paul trap offresonantly driving magnetic dipole transitions in the spin-rotation structure. The observation of the QLS signal of $L^+ = 2$ over \qty{19}{\hour} indicates that the rotational state is not decaying due to e.g.\ collisions with residual gas.}
\end{figure}

To demonstrate the capability of loading \Htwop in excited rotational states, we load a \Htwop ion with the REMPI laser tuned to the Q(2) resonance and probe characteristic transitions in its spin-rotation and Zeeman structure using QLS. See Appendix \ref{sec:spin-rot structure} for the spin-rotation and Zeeman structure and the probed transitions. For simplicity no spin-rotation-Zeeman state preparation is performed and only the spontaneously appearing QLS signal is observed, using the same protocol as in Sec. IV of Ref.~\cite{holzapfel2025quantum}. Examples of the QLS signal are shown in Fig. \ref{fig:qls}. We observe QLS signals over a duration of \qty{19}{\hour}. Similar to our experiments for $L^+=1$ in Ref.~\cite{holzapfel2025quantum}, the QLS signal disappears and reappears spontaneously. Here, we interpret this as changes in the spin-rotation-Zeeman state, driven not by collisions but by a different, thus far unknown disturbance. Potential causes are a weak ac magnetic field from the trap rf at \MHz{78.5} which offresonantly (detuning of about \SI{20}{\mega\hertz}) drives magnetic dipole transitions in the spin-rotation structure and/or the QLS operations leading to leakage to other spin-rotation-Zeeman states. 

In other molecular ion QLS experiments, the loss of the QLS signal has be attributed to rotational-state changes caused by thermal radiation \cite{chaffee2025high} and Raman scattering \cite{sinhal2020quantum}. These mechanisms are unlikely for our setup. Pure rovibrational transitions are dipole forbidden for homonuclear diatomic molecules such as \HtwopNoSpace. Raman scattering should be suppressed due to the large detuning and should populate also other rovibrational states. This would cause diffusion over the rovibrational manifold with a low probability of rescattering into the measured state and thus a low probability of the QLS signal reappearing. Chaffee \etal \cite{chaffee2025high} observed a dependence of the rotational-state loss rate on the experimental duty cycle, potentially indicating off-resonant effects from the QLS operations and motivating a similar study with our system.

The state changes are unlikely to occur due to collisions with residual gas. Collisions may lead to quenching of $L^+=2$ to $L^+=0$, but due to the low temperature of the residual gas and the large rotational splitting, a reexcitation from $L^+ = 0$ to $L^+=2$ due to collisions is extremely unlikely; assuming a residual gas temperature of \qty{10}{\kelvin} and using the $L^+=0,2$ splitting of \THz{5.2} \cite{hunter1974rotation}, we calculate an average population of \qty{2e-11} for the $L^+=2$ state. A deexcitation from $L^+=2$ would cause the QLS signal to disappear but not reappear, which does not match the signal in Fig.~\ref{fig:qls}. It is thus not plausible that the QLS signal blinking is due to changes of the rotational state.

\section{Conclusion and Outlook}
In this paper, we demonstrated loading of single \Htwop ions in an ion trap using (2+1) REMPI of residual \Htwo gas in the vacuum chamber. Our method provides high rotational selectivity, access to excited rotational states ($\nu^+ = 0$, $L^+ = 0, 1, 2$), and therefore isomer selectivity. It allowed us to observe transitions in the spin-rotation structure of the exited rotational state ($\nu^+ = 0$, $L^+ = 2$) by QLS.

In principle, loading of molecules with excited vibration, e.g. $\nu^+ = 1$, is also possible by tuning the REMPI wavelength to excite the vibrational state of \Htwo with two-photon absorption, e.g.\ from $\nu'' = 0$ to $\nu' = 3$, which gets predominantly photoionized to $\nu^+ = 1$~\cite{anderson1984resonance, fernandez2007dissociative}. However, the lower ionization probability per pulse and the high photodissociation probability for the ions make this challenging (see Appendix \ref{sec:simulation}). Loading $\nu^+ = 1$ molecules could be attempted by increasing the distance of the REMPI laser from the trap center to suppress the photodissociation further. This misalignment could lower the capture probability since the ion is now produced further from the trap center. One could potentially enhance the capture probability using a \BeNoSpace crystal instead of a single a \Be ion, which would provide stronger sympathetic cooling~\cite{tong2010}.

REMPI of background gas can serve as an efficient state-selective loading mechanism for novel single-ion high-precision spectroscopy of hydrogen molecular ions \cite{wellers2021controlled,konig2025nondestructive,holzapfel2025quantum}. The use of a low-density source, as opposed to a molecular beam, could be especially useful for the radioactive isotopologues $\mathrm{HT^+}$, $\mathrm{DT^+}$, and \TtwopNoSpace, which are of interest for the determination of the triton charge radius~\cite{bekbaev2013hyperfine, karr2025determination}.
The recently demonstrated control of single \HDpNoSpace molecules in a cryogenic Penning trap could be combined with the REMPI scheme demonstrated here to enable similar control over the homonuclear isotopologues~\cite{konig2025nondestructive}.

\begin{acknowledgments}
The authors thank F.~Merkt and R.~Hahn for stimulating discussions; N.~Schwegler for contributions to the experimental apparatus; A.~Ferk, M.~Stadler, and B.~Dönmez for control system and digital lock box support; and J.~P.~Home for useful discussions and continuing support. This work was supported by Swiss National Science Foundation Grant No.~212641, ETHZ Research Grant No. ETH-52 19-2, and an ETHZ Quantum Center QTNet grant. F.S. acknowledges financial support via an SNSF Swiss Postdoctoral Fellowship (Grant No.~224439). 
\end{acknowledgments}

\section*{Data Availability}
The data that support the findings of this study are available in the ETH Zürich Research Collection repository~\cite{kim2026state}.

\appendix

\section{Simulation of REMPI Excitation Dynamics}\label{sec:simulation}
We simulate the excitation dynamics of the (2+1) REMPI using a modified set of optical Bloch equations~\cite{dixit1985theory, haas2006two}. The equations used for the simulation are as follows:
\begin{subequations}\label{eq:obe}
\begin{align}
    \frac{\partial}{\partial t}\rho_{gg} &= - \Omega\, \mathrm{Im}(\rho_{ge}) + \Gamma_s \rho_{ee}, \\
    \frac{\partial}{\partial t}\rho_{ge} &= 
    i\frac{\Omega}{2}(\rho_{gg} - \rho_{ee})
    - \left(\frac{\Gamma_i + \Gamma_s+\Gamma_L}{2}\right) \rho_{ge},\\
    \frac{\partial}{\partial t}\rho_{ee} &= \Omega\, \mathrm{Im}(\rho_{ge}) - (\Gamma_i + \Gamma_s) \rho_{ee}, \\
    \frac{\partial}{\partial t}\rho_{ii} &= \alpha\Gamma_i\rho_{ee} - \Gamma_d\rho_{ii},
\end{align}
\end{subequations}
where $\rho$ is the density matrix inside the three-level Hilbert space spanned by the ground state \XState ($g$), intermediate state \EState ($e$), and ionic ground state $X\;^2\Sigma_g^+$ ($i$).

First, we simulate the ionization with vibrational levels $\nu''=0 \rightarrow \nu'=0 \rightarrow \nu^+=0$. The spontaneous decay rate $\Gamma_s = \SI{5e6}{\second^{-1}}$ and total ionization cross section $\sigma_i = \SI{3e-18}{\centi\meter^2}$ of the \EState state are given in Refs.~\cite{chandler1986measured, fernandez2007dissociative}, respectively. The vibrational branching ratio of the ionization step is denoted $\alpha$. At \SI{202}{\nano\meter}, $\alpha$ is $\sim 0.9$ for $\nu'=0\rightarrow\nu^+=0$~\cite{anderson1984resonance, rudolph19872+, cornaggia1987photo}. The total ionization rate $\Gamma_i$ is then calculated using $\Gamma_i = \sigma_i (I/h \nu)$, where $I$ and $h \nu$ are the intensity and the photon energy of the REMPI laser. The ionization rate for $\nu^+=0$ is $\alpha\Gamma_i$. The two-photon transition moment of 9.12~a.u.\ for the \XState ($\nu''=0$, $L''=0$) $\rightarrow$ \EState ($\nu'=0$, $L'$=0) state from Ref.~\cite{pomerantz2004line} can be used with the REMPI beam intensity to calculate the effective Rabi frequency $\Omega$ between states $g$ and $e$~\cite{haas2006two}. The widths of the resonances in Fig.~\ref{fig:two_photon} correspond to a linewidth of approximately $\Gamma_L = 2 \pi \times \qty{400}{GHz}$ at \qty{101}{nm}. We take this into account by adding $\Gamma_L/2$ to the dampening of the coherence term in Eq.~\hyperref[eq:obe]{A1b}, following Ref.~\cite{von2020theory}. We assume that the two-photon excitation from the \XState state to the \EState state is resonant in our simulations.

We also add a decay of the ionized state population due to photodissociation in Eq.~\hyperref[eq:obe]{A1d}, where the dissociation rate $\Gamma_d = \sigma_d(I/\hbar\omega)$ can be calculated using the photodissociation cross sections $\sigma_d$ found in Ref.~\cite{dunn1968photo, dunn1968tables}. At \nm{202}, the photodissociation cross section is \qty{1.9e-20}{\centi\meter^2} for $\nu^+ = 0$ and \qty{9.0e-19}{\centi\meter^2} for $\nu^+ = 1$.

\begin{figure}
\includegraphics[width=0.48\textwidth]{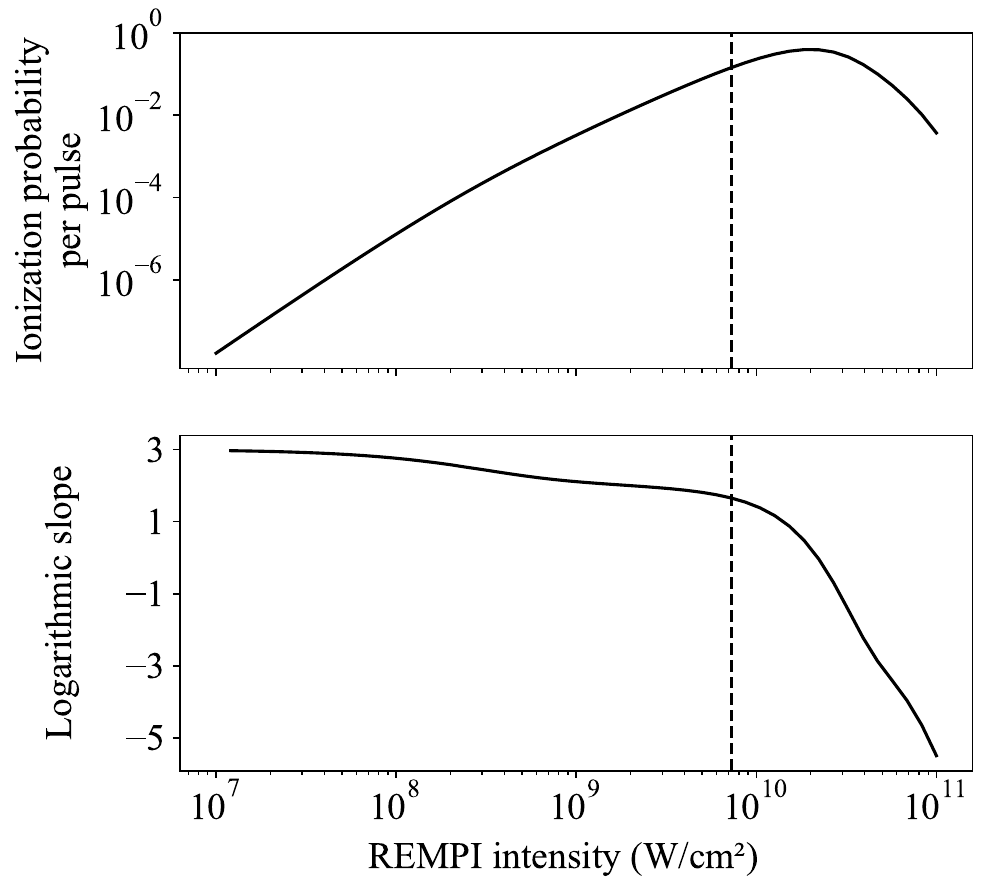}
\caption{\label{fig:obe} \textbf{Simulated ionization probability and power-law dependence.} Ionization probability per pulse for $(\nu^+ = 0$, $L^+=0$) is plotted against varying REMPI beam intensities. The expected spatially averaged peak intensity in our setup of \SI{7}{\giga\watt/\centi\meter^2} is indicated by the dashed vertical lines. The simulated ionization probability at this intensity is \SI{20}{\percent}. The ionization probability reaches its maximum of \SI{40}{\percent} at \SI{18}{\giga\watt/\centi\meter^2}, and starts to decline for larger intensities due to photodissociation. The bottom plot shows the logarithmic slope of the top plot. At low intensities, we extract a power law exponent of $\beta = 3$ as expected. At \SI{7}{\giga\watt/\centi\meter^2}, we predict an exponent of $\beta = 1.4$.}
\end{figure}

Equations~\hyperref[eq:obe]{A1a-d} were numerically integrated for a pulse duration of \SI{3}{\nano\second} to calculate the ionization probability per pulse, for varying intensities. We use the spatially averaged peak intensity as defined in Sec.~\ref{sec:exp_setup}, as opposed to spatial peak intensity, to simulate the mean ionization dynamics within the beam waist. The results for $\nu^+ = 0$ are shown in Fig. \ref{fig:obe}. The ionization probability for our REMPI intensity is simulated to be \SI{20}{\percent}. The probability increases with intensity to a maximum of \SI{40}{\percent} at \SI{18}{\giga\watt/\centi\meter^2} and then starts to decrease due to photodissociation. For the power law dependence $I^\beta$, we extract $\beta = 3$ at low intensities, which is expected since (2+1) REMPI is a three-photon process. As the REMPI intensity increases, the logarithmic slope decreases due to saturation. At our typical REMPI intensity, we extract $\beta = 1.4$. 

We also repeat the simulation with $\alpha = 0.1$ and the photodissociation cross section $\sigma_d$ updated for $\nu^+ = 1$ to estimate the vibrational selectivity of the ionization. At our REMPI intensity, we expect \qty{0.1}{\percent} ionization probability per pulse for $\nu^+=1$. Thus, for an \Htwop ion loaded after a single REMPI pulse, the probability of the ion being in vibrational state $\nu^+=0$ is \qty{99.5}{\percent}. 

We then simulate the ionization with vibrational levels $\nu''=0 \rightarrow \nu'=3 \rightarrow \nu^+=1$ to investigate the feasibility of loading an $\nu^+=1$ ion. The total ionization cross section from $\nu'=3$ is also $\sigma_i = \SI{3e-18}{\centi\meter^2}$~\cite{fernandez2007dissociative}, while the vibrational branching ratio $\alpha$ of the ionization step is $\approx 0.6$~\cite{anderson1984resonance}. The two-photon transition moment of 9.12~a.u. from the previous simulation is multiplied by the ratio of the Franck-Condon overlaps of 0.2148 between vibrational states $\nu''=0$ and $\nu'=3$ and 0.1579 between $\nu''=0$ and $\nu'=0$~\cite{ritchie1982three}.

At our REMPI intensity, we expect the ionization probability per pulse to be \qty{1}{\percent} for $\nu^+=1$ production, as opposed to \qty{20}{\percent} for $\nu^+=0$. Furthermore, after an ion with $\nu^+ = 1$ is loaded, the photodissociation probability (assuming a distance of \um{40} between the REMPI beam and the ion) is \qty{40}{\percent}, as opposed to \qty{1}{\percent}. Therefore, we do not expect any significant speedup from sending bursts of pulses, which it more difficult to load an ion in $\nu^+ = 1$.

\section{H$_2^+$ ($\mathbf{\nu^+ = 0, L^+ = 2}$) spin-rotation and Zeeman structure}\label{sec:spin-rot structure}
\begin{figure}
\includegraphics[width=0.48\textwidth]{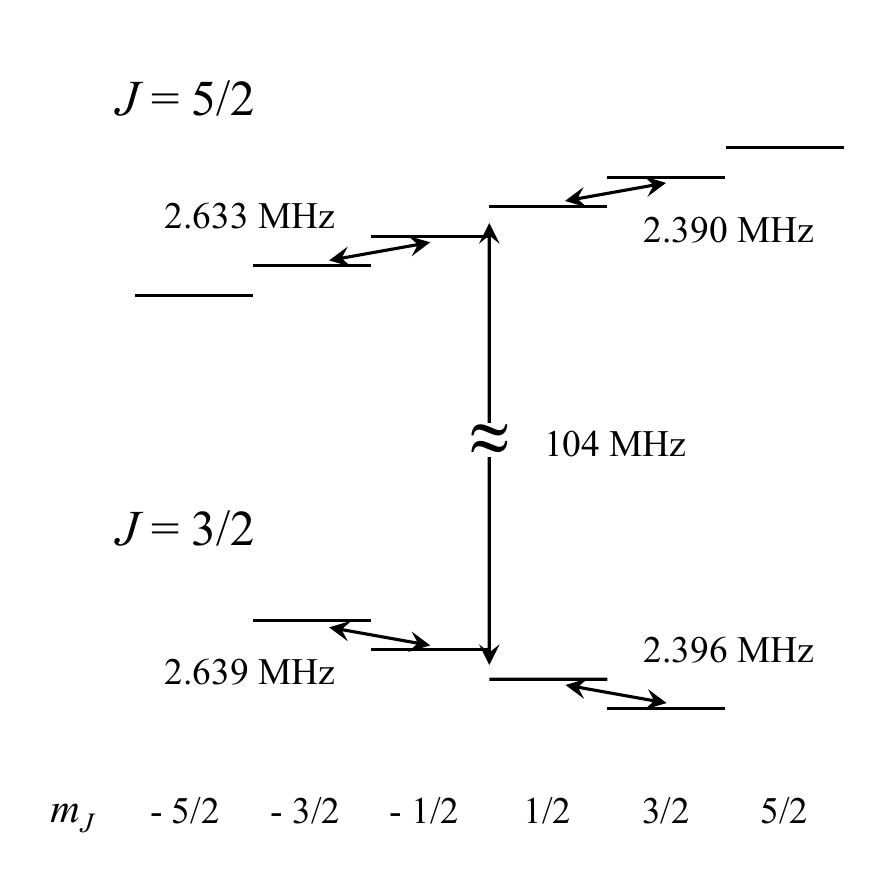}
\caption{\label{fig:spin-rot structure} \textbf{Spin-rotation and Zeeman structure of the \Htwop $\nu^+ = 0, L^+ = 2$ state.} The energy level structure is given for the magnetic bias field applied in the experiment (\qty{450.135}{\micro\tesla}). The spin-rotation splitting is calculated to be \MHz{104}. The observed QLS signals are consistent with the indicated transitions. The deviation of the resonance frequencies from the theoretical predictions can be explained by ac Stark shifts due to the $1050$-$\mathrm{nm}$ probe lasers.}
\end{figure}

Fig. \ref{fig:spin-rot structure} shows the spin-rotation and Zeeman structure of the \Htwop ($\nu^+ = 0$, $L^+ = 2$) state for our magnetic bias field of \qty{450.135}{\micro\tesla}. QLS signals were observed at 2.650(25), 2.645(25), and \MHz{2.395(25)}. Taking into account the ac Stark shift induced by the $1050$-$\mathrm{nm}$ probe lasers of up to \kHz{30}, these transition values are in approximate agreement with the theoretical predictions: The transitions $\ket{J=3/2, m_J=-1/2}\leftrightarrow\ket{J=3/2, m_J=-3/2}$ (\MHz{2.639}) and $\ket{J=5/2, m_j=-1/2}\leftrightarrow\ket{J=5/2, m_J=-3/2}$ (\MHz{2.633}) likely match the observed values 2.650(25) and \MHz{2.645(25)}. Either of the transitions $\ket{J = 3/2, m_J = 1/2}\leftrightarrow\ket{J=3/2, m_J=3/2}$ (\MHz{2.396}) and $\ket{J=5/2, m_J=1/2}\leftrightarrow\ket{J=5/2, m_J=3/2}$ (\MHz{2.390}) could be the observed signal at \MHz{2.395(25)}.

\section{Two-photon rotational line strengths and rotational selectivity of the ionization step}\label{sec:theory}
The first step in (2+1) REMPI is the two-photon excitation of the molecular ground state to the \EState intermediate state. It turns out that the Q transitions ($\Delta L = 0$) are much stronger than the O ($\Delta L = -2$) and S ($\Delta L = +2$) transitions. For $\Sigma \rightarrow \Sigma$ transitions with linear polarization, the rotational line strengths are given by~\cite{bray1975rovibronic}
\begin{subequations}\label{eq:line_strength} 
\begin{align}
    R_{\mathrm{O}(L)} &= \frac{L(L-1)}{30(2L-1)}\mu_S^2,\\
    R_{\mathrm{Q}(L)} &= \frac{2L+1}{9}\mu_I^2 + \frac{L(L+1)(2L+1)}{45(2L-1)(2L+3)}\mu_S^2,\\
    R_{\mathrm{S}(L)} &= \frac{(L+1)(L+2)}{30(2L+3)}\mu_S^2,
\end{align}
\end{subequations}
where $\mu_I$ and $\mu_S$ are ``trace scattering'' and ``zero trace symmetric tensor scattering'' terms arising from vibronic contributions~\cite{marinero1983the}. Substituting $\mu_I^2/\mu_S^2$ of 5.49 extracted from Ref.~\cite{marinero1983the}, we expect the O and S transitions to be suppressed relative to Q transitions by a factor of more than 27.

For molecules with isotropic rotational motion, the angular distribution of the ejected photoelectron during ionization can be described by the average anisotropy parameter~\cite{cooper1968angular}
\begin{equation}
\bar{\beta}(L') = \frac{\sum_{L^+}\sigma_{L'\rightarrow L^+}\beta_{L'\rightarrow L^+}}{\sum_{L^+}\sigma_{L'\rightarrow L^+}},
\end{equation}
where $\sigma_{L'\rightarrow L^+}$ and $\beta_{L'\rightarrow L^+}$ are the cross sections and anisotropy parameters for different ionization channels leading to the ionic rotational quantum number $L^+$~\cite{dill1972angular}, respectively. Assuming that the anisotropy of the \EState state is dominated by $d$ character~\cite{sichel1970angular}, we can truncate the average to three terms; $L^+=L'$ and $L^+=L'\pm 2$. For $L' = 0$, only two terms remain, and therefore we can estimate the rotational selectivity of $\bar{\beta}$ by solving for $\sigma_{L'\rightarrow L^+=L'} / \sum_{L^+}\sigma_{L'\rightarrow L^+}$~\cite{cornaggia1987photo}. Reference ~\cite{perreault2016angular} calculated $\bar{\beta}(L'=0)$ of 1.72 for the same \Htwo REMPI scheme (with $\nu' = 0$) using a time-of-flight measurement. From $\beta_{0\rightarrow 0} = 2$  and $\beta_{0\rightarrow 2} = 1/5$~\cite{dill1972angular}, we estimate that $L^+$ will stay unchanged by ionization following the Q(0) transition with \qty{84}{\percent} probability. 

To the best of our knowledge, no studies directly compute the anisotropy parameter (or rotational selectivity) of ionization from the \EState $(\nu'=0, L'=1)$ state. Photoelectron angular distribution from (n+1) REMPI of $L'\neq 0$ should in principle be different from one from REMPI of $L'=0$ due to the change in molecular alignment during multiphoton absorption~\cite{dixit1985theory}. However, studies using \Htwo in $\nu'=1$~\cite{anderson1984resonance} and $\nu'=3$~\cite{rudolph19872+} reported $\bar{\beta}(L'=0) \approx \bar{\beta}(L'=1)$ for ejected photoelectrons after Q(0) and Q(1) transitions, respectively, which suggests that the effect of molecular alignment on the anisotropy parameter and therefore the rotational selectivity is minimal.

\bibliography{h2_rempi}
\newpage
\clearpage
\pagebreak

\renewcommand{\bibliography}[1]{} 

\end{document}